\documentclass[sigconf]{acmart}
\usepackage{multirow}
\usepackage[table]{xcolor}
\usepackage{pifont}
\usepackage[switch]{lineno}  % 支持双栏行号
\usepackage{etoolbox}
\usepackage[table]{xcolor}
\usepackage[normalem]{ulem}
\usepackage{booktabs}
\usepackage{multirow}
\usepackage{array}
\usepackage{makecell}
\usepackage{graphicx}
\usepackage{threeparttable}
\usepackage{bm}
\definecolor{cmp}{HTML}{E8F5E9}
\definecolor{base}{HTML}{E3F2FD}
\definecolor{abl}{HTML}{E8EAF6}
\definecolor{enc}{HTML}{EDE7F6}
\definecolor{rob}{HTML}{FFF3E0}

\definecolor{RColor}{HTML}{0805CD}
\AtBeginDocument{%
  }

\setcopyright{acmlicensed}
\copyrightyear{2026}
\acmYear{2026}
\setcopyright{cc}
\setcctype{by}
\acmConference[MM '26] {Proceedings of the 34th ACM International Conference on Multimedia}{November 10--14, 2026}{Rio de Janeiro, Brazil.}
\acmBooktitle{Proceedings of the 34th ACM International Conference on Multimedia (MM '26), November 10--14, 2026, Rio de Janeiro, Brazil}
\acmDOI{10.1145/3767308.3835861}
\acmISBN{979-8-4007-2213-4/2026/11}
\begin{document}

\title{Time-Frequency Consistency Learning for Robust Speech Deepfake Detection }
\author{Jun Xue}
\email{junxue@whu.edu.cn}
\orcid{0009-0001-8465-011X}
%\email{webmaster@marysville-ohio.com}
\affiliation{%
	\institution{School of Cyber Science and Engineering, Wuhan University}
	\city{Wuhan}
	\country{China}
}

\author{Zhuolin Yi}
\email{yizhuolin@whu.edu.cn}
\orcid{0009-0008-5771-8976}
\affiliation{%
	\institution{School of Cyber Science and Engineering, Wuhan University}
	\city{Wuhan}
	\country{China}
}

\author{Yanzhen Ren}
\authornote{Corresponding author.} % 这一行是添加的标记
\email{renyz@whu.edu.cn}
\orcid{0000-0003-0799-5082}
\affiliation{%
	\institution{School of Cyber Science and Engineering, Wuhan University}
	\city{Wuhan}
	\country{China}
}
\author{Yihuan Huang}
\email{yihuanhuang@whu.edu.cn}
\orcid{0009-0000-8939-0209}
\affiliation{%
		\institution{Wuhan University}
	\city{Wuhan}
	\country{China}
}
\author{Jiayu Xiong}
\email{yuinst@outlook.com}
\orcid{0009-0000-3836-3438}
\affiliation{%
	\institution{Tongji University}
	\city{Shanghai}U
	\country{China}
}
\author{Yi Chai}
\email{talent@whu.edu.cn}
\orcid{0009-0006-7553-974X}
\affiliation{%
		\institution{Wuhan University}
	\city{Wuhan}
	\country{China}
}

\author{Guanxiang Feng}
\email{2022302191328@whu.edu.cn}
\orcid{0009-0001-6742-1702}
\affiliation{%
%	\institution{School of Cyber Science and Engineering, Wuhan University}
		\institution{Wuhan University}
	\city{Wuhan}
	\country{China}
}
\author{Jiajun Liu}
\email{jiajunliu@whu.edu.cn}
\orcid{0009-0006-7070-9172}
\affiliation{%
		\institution{Wuhan University}
	\city{Wuhan}
	\country{China}
}
\author{Tong Zhang}
\email{toby\_zt@outlook.com}
%\orcid{0009-0006-7070-9172}
\affiliation{%
		\institution{Wuhan University}
	\city{Wuhan}
	\country{China}
}

\renewcommand{\shortauthors}{Jun Xue et al.}

\begin{abstract}
Recently, speech deepfake detection (SDD) has achieved significant progress. However, its robustness evaluation remains largely confined to controlled additive noise scenarios, lacking systematic investigation of the complex distortions introduced by acoustic front-end (AFE) processing pipelines in real-world deployments. In this work, we simulate a unified AFE pipeline comprising acoustic echo cancellation, noise suppression, automatic gain control, and voice activity detection (VAD), and conduct a comprehensive evaluation of current state-of-the-art models. The results show that the nonlinear and time-frequency coupled distortions introduced by AFE significantly degrade detection performance.
To address this issue, we propose a Time-Frequency Consistency Learning (TFCL) framework, which aims to learn invariant spoofing representations that remain stable before and after AFE processing. We observe that AFE not only introduces temporal misalignment (e.g., segment-level shifts caused by VAD), but also weakens or distorts critical frequency-domain cues. To this end, TFCL employs an attention-driven soft alignment mechanism to capture cross-temporal dependencies, along with frequency-domain structural consistency constraints to enforce feature invariance. As a result, the model is able to maintain stable representations under both temporal perturbations and spectral distortions.
Extensive experimental results demonstrate that the proposed method effectively mitigates the performance degradation caused by AFE processing, significantly improving the robustness of SDD in real-world scenarios. The code is available at \url{https://github.com/JunXue-tech/TFCL}.
\end{abstract}


% \includegraphics[width=\textwidth]{sampleteaser}


\maketitle

\section{Introduction}
\label{sec:intro}
With the advancement of generative models, text-to-speech (TTS)  \cite{zhou2025voxcpm}  and voice conversion (VC) \cite{du2024cosyvoice} techniques are now capable of producing highly realistic speech signals, enabling speech deepfakes to be used in real-world attacks. While these technologies offer significant benefits, they have also been exploited for fraudulent activities, identity impersonation, and other malicious purposes, posing substantial risks to individual security and social trust. In this context, speech deepfake detection (SDD) has emerged as an important research topic. In recent years, benchmark initiatives such as ASVspoof \cite{todisco2019asvspoof,liu2023asvspoof,wang2024asvspoof} and ADD \cite{yi2022add,yi2023add} have driven the development of detection methods, leading to continuous performance improvements under controlled conditions.

To improve the applicability of SDD models in real-world scenarios, existing studies have begun to investigate robustness under noisy conditions \cite{cao2025robust,chen2025adaptive}. For instance, ADD 2022 \cite{yi2022add} introduces real-world environmental noise and background music to evaluate model performance under complex acoustic conditions. At the methodological level, Fan et al. \cite{fan2024dual} enhance robustness in noisy scenarios through a dual-branch distillation strategy, while Sen et al.  \cite{sen2025toward} systematically analyze performance variations across different signal-to-noise ratio (SNR) levels, revealing the impact of environmental noise on detection performance.
However, these works primarily model speech degradation as additive noise, overlooking the complex signal processing pipeline that noisy speech undergoes in real-world communication systems.

In practical applications, speech deepfake threats have expanded to real-time communication (RTC) platforms~\cite{xue2026rtcfake}, with their latent risks manifesting in high-profile security incidents. A prime example \cite{cna_deepfake_scam_2025} is a 2025 case where attackers utilized AI-generated voices during a Zoom meeting to successfully impersonate a corporate CEO, executing a \$499,000 fraud. Such incidents underscore the necessity for Speech Deepfake Detection (SDD) methods to maintain high reliability in authentic communication deployments. However, within mainstream communication applications, speech signals are not transmitted directly but are processed by an acoustic front-end (AFE) pipeline, including acoustic echo cancellation (AEC), noise suppression (NS), automatic gain control (AGC), and voice activity detection (VAD). This pipeline introduces complex nonlinear transformations due to the cascading of multiple modules. As a result, critical acoustic artifacts of spoofed speech may be distorted or suppressed. Unfortunately, existing research lacks a thorough investigation of this issue.

To address these challenges, we first simulate the acoustic front-end processing and conduct a systematic evaluation of state-of-the-art detection models. Experimental results reveal a severe performance degradation across existing models following AFE processing. By further analyzing the underlying impact of AFE on speech signals, we observe two critical phenomena: in the time domain, VAD and AGC introduce segment-level clipping and nonlinear gain variations, leading to significant temporal misalignment; in the frequency domain, NS and AEC inevitably suppress or distort essential spectral cues while mitigating interference.

Motivated by these findings, we propose a Time-Frequency Consistency Learning (TFCL) framework, designed to learn invariant forgery representations across the AFE pipeline. Specifically, TFCL addresses two types of AFE-induced distortions. In the time domain, an attention-driven soft alignment mechanism is employed to mitigate temporal inconsistencies by modeling cross-temporal dependencies. In the frequency domain, we introduce structural consistency constraints to preserve stable discriminative representations under spectral distortions. By jointly enforcing temporal dependency alignment and frequency structural consistency, the proposed framework is able to learn robust representations that remain stable before and after AFE processing. Extensive experimental results demonstrate that the proposed method significantly enhances the robustness of SDD models against distortions introduced by AFE processing chains.

The main contributions of this work are summarized as follows:

\begin{itemize}
	\item We move the robustness evaluation of SDD beyond additive-noise settings to realistic AFE distortions, and demonstrate their significant impact on detection performance.
	
	\item We show that AFE degradation is not a single corruption type, but can be characterized as a coupled shift in temporal dependency and frequency structure, which challenges the stability of spoofing representations.
	
	\item We propose a time-frequency consistency learning framework that addresses these two shifts through temporal soft alignment and frequency structural consistency regularization, yielding consistent improvements across different AFE conditions and model backbones.
\end{itemize}

%\vspace{-2ex}
\section{Related Work}

\textbf{Speech Deepfake Detection Datasets.} In recent years, a variety of datasets and benchmarks have been developed for SDD, primarily focusing on robustness from the perspectives of input degradation and transmission processes. On the one hand, datasets such as ADD 2022~\cite{yi2022add} introduce real-world environmental noise and background music to evaluate model robustness under low SNR and complex acoustic conditions. On the other hand, some studies further consider factors in the communication pipeline, including speech codecs~\cite{liu2023asvspoof}, compression artifacts~\cite{wang2024asvspoof}, and bandwidth constraints~\cite{wang2024asvspoof}, and even extend to neural codec~\cite{xie2025codecfake,chen25j_interspeech} scenarios to simulate realistic speech transmission in modern communication systems. These efforts have systematically expanded the evaluation scope of SDD at the data level. However, existing work mainly focuses on input signal degradation or transmission compression, and lacks systematic investigation and modeling of the complex distortions introduced by AFE processing pipelines, which are widely present in RTC systems.

\noindent\textbf{Speech Deepfake Detection Methods.}
Mainstream SDD systems typically combine a front-end feature extractor with a back-end classifier. Wav2vec 2.0-based representations~\cite{baevski2020wav2vec} have been widely adopted for modeling spoofing artifacts~\cite{xue2026profiling}, while back-end architectures such as AASIST~\cite{jung2022aasist}, Mamba~\cite{chen2024rawbmamba}, and Nes2Net~\cite{liu2025nes2net} improve detection through enhanced structural and temporal modeling. Robustness has also been explored through training strategies. For example, DBKD~\cite{fan2024dual} employs knowledge distillation to improve stability under noisy conditions, whereas RawBoost~\cite{tak2022rawboost} simulates signal degradation to enhance robustness against codec and channel variations.

Nevertheless, existing methods mainly target additive noise or transmission-related distortions, with limited attention to AFE processing in real-world communication systems. Unlike a single corruption, AFE comprises cascaded modules that introduce accumulated nonlinear and time-frequency coupled distortions, making spoof-discriminative representations difficult to preserve. To address this issue, we construct a unified AFE simulation pipeline, systematically evaluate its impact on existing SDD models, and propose a TFCL framework to mitigate temporal dependency disruption and frequency structural distortion.

\begin{figure}[!t]  % !t = 强制放在页面顶部
	\centering
	\includegraphics[scale=0.3]{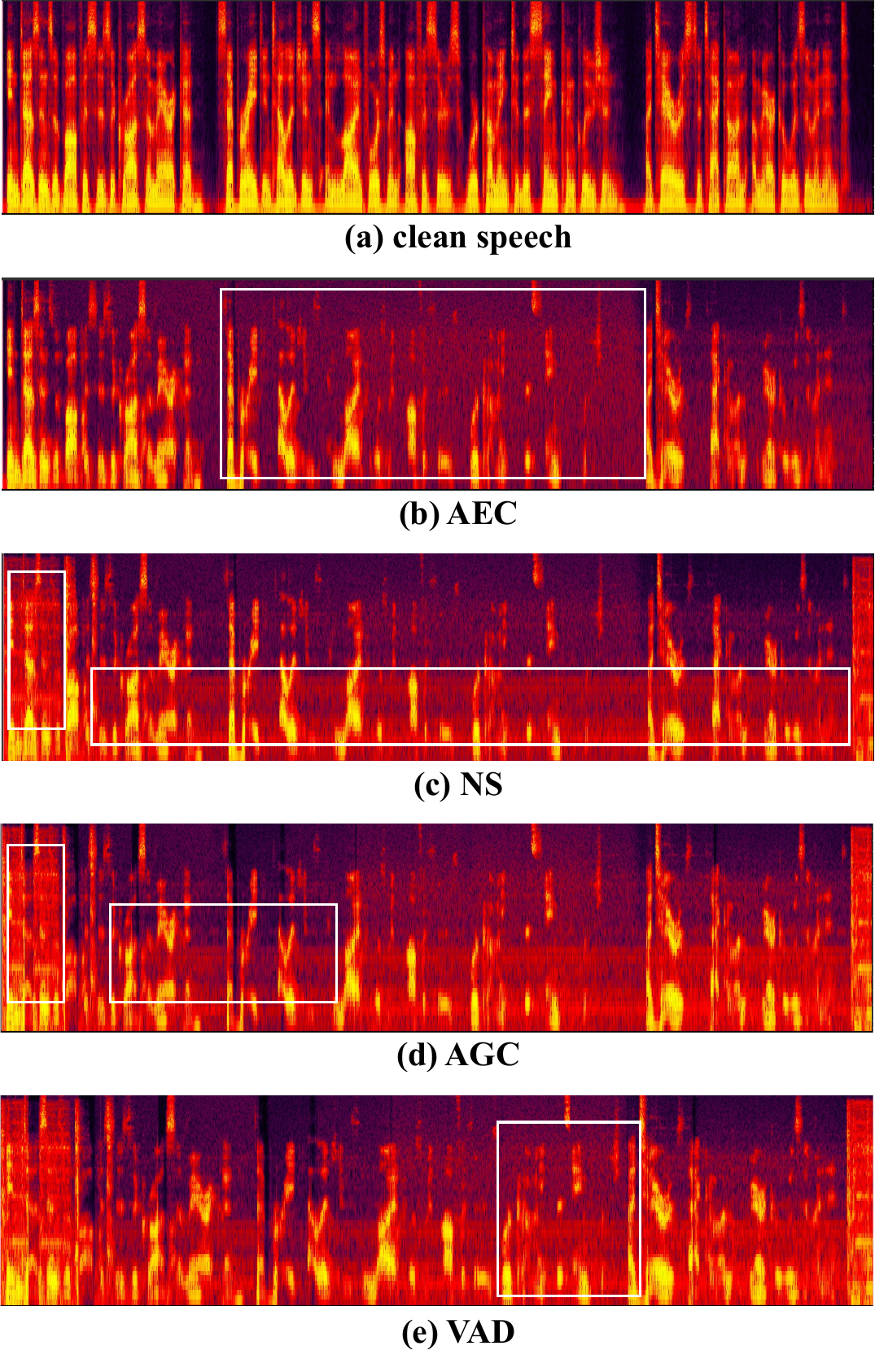}
%	\vspace{-2ex}
	\caption{Spectrogram visualization of the utterance LA\_E\_1184038.wav across different front-end processing stages. The detailed processing pipeline is described in Section~\ref{subsec:sim}.}
	\label{fig:spec}
\end{figure}
\section{Methodology}

\subsection{Problem Formulation}

Given a clean speech signal $x_c$ and its corresponding distorted version $x_d = \mathcal{A}(x_c)$ processed by the AFE pipeline $\mathcal{A}(\cdot)$, our goal is not to enforce strict signal-level or frame-wise equivalence between $x_c$ and $x_d$. Instead, we aim to learn \emph{AFE-invariant spoofing representations} that preserve spoof-discriminative semantics while being robust to AFE-induced nuisance variations.

Specifically, we focus on enforcing \emph{discriminative representation consistency} between clean and distorted speech, rather than waveform identity or exact temporal alignment. This formulation allows the model to maintain stable spoofing cues under complex front-end processing distortions.

\subsection{Acoustic Front-end Pipeline}

In modern real-time speech communication systems (e.g., VoIP, video conferencing, and WebRTC-based applications), the acoustic front-end typically consists of a cascade of speech enhancement modules, including AEC, NS, AGC, and VAD. This modular architecture has become a de facto standard in industrial systems. For instance, the Audio Processing Module (APM) in WebRTC explicitly integrates AEC, NS, and AGC as core components for real-time processing of microphone signals \cite{webrtc_apm}. These processing modules are generally applied after audio capture and before encoding, forming a complete acoustic front-end processing pipeline \cite{webrtc_pipeline}.

From a processing perspective, mainstream systems typically organize these modules in the order of “AEC → NS → AGC → VAD”. This ordering is motivated by the dependency among different types of distortions: acoustic echo is a structured interference that is highly correlated with the far-end signal and therefore must be removed first; background noise is largely uncorrelated and its suppression benefits from prior echo reduction; AGC is then applied to normalize signal amplitude for stable downstream processing; finally, VAD operates on relatively clean and amplitude-stabilized signals to detect speech activity. In communication systems, VAD is further used to identify active speech segments, thereby avoiding redundant encoding and transmission of silent intervals and improving bandwidth efficiency. Similar module configurations and processing paradigms have been widely adopted in real-time speech communication research \cite{wang2022nn3a}.

\subsection{Analysis and Motivation}

\begin{figure*}[!t]  % !t = 强制放在页面顶部
	\centering
	\includegraphics[scale=0.55]{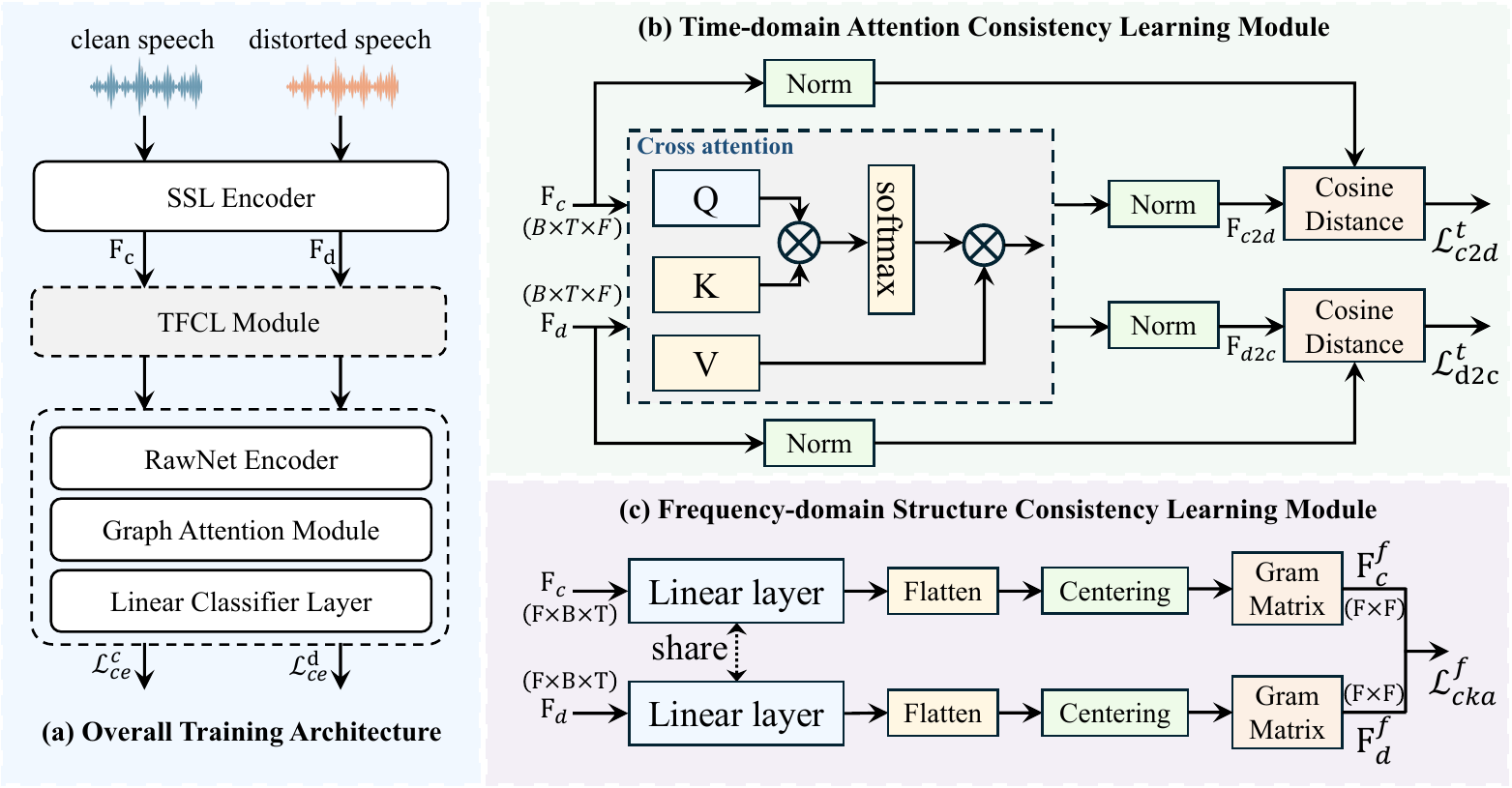}
	\caption{Overview of the proposed framework. The distorted speech is generated by passing clean speech through the AFE pipeline. The backend classifier adopts the AASIST architecture. \textbf{\textit{Note:}} The dual-input design and the TFCL module are used only during training for consistency learning, and are removed during inference without introducing additional computational overhead.} 
	\label{fig:training}
\end{figure*}

From the spectrogram visualization in Figs.~\ref{fig:spec}, it can be observed that due to distortion accumulation introduced by the speech front-end processing pipeline, the spectral characteristics and temporal prosody of the original speech signal are irreversibly degraded. This degradation manifests in different forms across each processing stage:

In the AEC stage, although echo components are effectively suppressed, it simultaneously introduces substantial semantic information loss. As observed in the highlighted regions, portions of continuous speech energy are significantly attenuated or even completely removed, resulting in the emergence of extended silent segments. Moreover, AEC fails to fully eliminate background interference, leaving a noticeable level of residual noise. This combination of “over-suppression and residual noise” introduces latent risks for subsequent processing stages.

The NS stage effectively suppresses high-frequency background noise, resulting in a cleaner spectral structure in the high-frequency regions of the spectrogram. However, this process also introduces a certain degree of spectral variation. As observed in Fig.~\ref{fig:spec} (c), parts of the spectral structure exhibit mild distortion, manifested as reduced harmonic continuity and uneven local energy distribution. In addition, noticeable residual noise remains in the low-frequency region.

Subsequently, AGC adjusts the gain to constrain the speech signal within a certain dynamic range. However, since distortions have already been introduced in the preceding stages, AGC may further amplify these artifacts. On the one hand, previously attenuated or already distorted spectral components can be nonlinearly enhanced under dynamic gain adjustment, making the distortions more pronounced. On the other hand, the energy distribution across different temporal segments is rebalanced, thereby weakening the natural dynamic variations of speech and exacerbating temporal inconsistencies.

Finally, in the VAD stage, the system detects and segments speech activity regions based on energy or statistical features, and typically removes non-speech segments (e.g., silence) to reduce redundancy and lower transmission or storage costs. However, since preceding processing stages (e.g., AEC) may introduce non-natural silence or speech loss, VAD-based segmentation on such inputs can further alter the temporal structure of the speech, thereby introducing additional effects on speech rhythm (prosody) and overall prosodic patterns.

Overall, the above analysis indicates that the speech front-end processing pipeline introduces progressively accumulated perturbations to the intrinsic structure of speech in both temporal and frequency domains. These perturbations are not only reflected in local information loss or energy variation, but more importantly in the disruption of temporal dependencies and alterations of spectral structural relationships. Based on these observations, we identify two key representation-level effects introduced by AFE processing: (1) \textbf{temporal dependency disruption}, and (2) \textbf{frequency structural distortion}.
These two effects call for alignment at different granularities: in the time domain, relation-aware soft alignment is required to handle non-rigid temporal mismatch; in the frequency domain, global structural consistency is needed to preserve discriminative spectral relationships. Motivated by this decomposition, we propose a time-frequency consistency learning framework, as illustrated in Fig.~\ref{fig:training}, which jointly addresses temporal and frequency-domain distortions.

\subsection{Overview of the Proposed Framework}

Given clean and distorted speech, we first extract frame-level representations using a shared feature encoder, denoted as $\bm{F}_c$ and $\bm{F}_d$. To address the two types of AFE-induced distortions, we introduce a dual-branch consistency learning framework.

In the time domain, a temporal dependency consistency module is designed to mitigate temporal misalignment by modeling cross-temporal dependencies through soft alignment. In the frequency domain, a structural consistency module is introduced to preserve global spectral relationships under front-end processing.

The two consistency objectives are jointly optimized together with the classification loss, enabling the model to learn spoof-discriminative representations that remain stable before and after AFE processing.

\subsection{Temporal Dependency Consistency Learning}

Motivated by the temporal dependency disruption introduced by AFE processing (e.g., segment-level shifts and non-rigid temporal mismatch caused by VAD and AGC), we propose a temporal dependency consistency learning mechanism to align clean and distorted speech representations.

Specifically, clean and distorted speech signals are first processed by a shared feature extractor to obtain frame-level representations. Let $\bm{F} = [\bm{f}_1, \bm{f}_2, \dots, \bm{f}_T] \in \mathbb{R}^{T \times D}$ denote the sequence of frame-level features, where $T$ is the number of time steps and $D$ is the feature dimension. The corresponding representations for clean and distorted speech are denoted as $\bm{F}_c$ and $\bm{F}_d$, respectively.

Due to the non-rigid temporal distortions introduced by AFE processing, strict frame-wise alignment is not appropriate. Therefore, we employ a bidirectional cross-attention mechanism to perform relation-aware soft alignment between $\bm{F}_c$ and $\bm{F}_d$.

For notational convenience, let $(\bm{F}_s, \bm{F}_r) \in \{(\bm{F}_c, \bm{F}_d), (\bm{F}_d, \bm{F}_c)\}$ denote the source-target feature pair for each direction. The cross-attention aligned representation is then formulated as:
\begin{equation}
	\bm{F}_{s \rightarrow r}
	=
	\mathrm{Softmax}
	\left(
	\frac{\bm{Q}_s \bm{K}_r^{\top}}{\sqrt{D}}
	\right)
	\bm{V}_r,
\end{equation}

To enforce temporal consistency, we measure the discrepancy between the aligned feature and its corresponding original feature using cosine distance:
\begin{equation}
	\mathcal{L}_{s \rightarrow r}^{t}
	=
	1
	-
	\frac{
		\bm{F}_{s \rightarrow r} \cdot \bm{F}_s
	}{
		\|\bm{F}_{s \rightarrow r}\| \, \|\bm{F}_s\|
	}.
\end{equation}

The final temporal consistency loss is defined as:
\begin{equation}
	\mathcal{L}_{t}
	=
	\frac{1}{2}
	\left(
	\mathcal{L}_{c \rightarrow d}^{t}
	+
	\mathcal{L}_{d \rightarrow c}^{t}
	\right).
\end{equation}

The aligned representation aggregates supportive evidence from the counterpart domain, while preserving the source-domain spoof semantics, rather than collapsing the two domains into identical embeddings. This design allows the model to learn temporally robust representations under AFE-induced distortions.

\begin{figure}[!t]  % !t = 强制放在页面顶部
	\centering
	\includegraphics[scale=0.54]{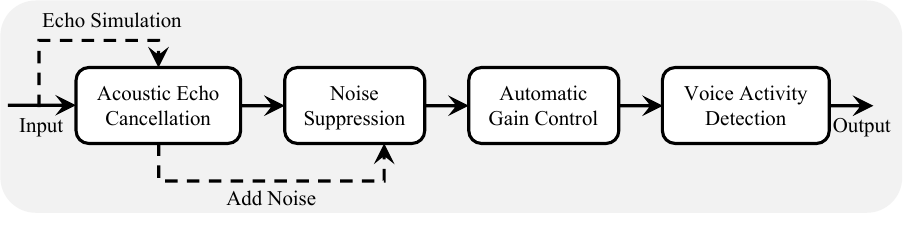}
	\caption{General acoustic front-end processing pipeline}
	\label{fig:afe}
\end{figure}

\subsection{Frequency Structural Consistency Learning}

Motivated by the frequency structural distortion introduced by AFE processing, we further propose a frequency structural consistency learning mechanism to preserve stable discriminative spectral relationships.

Unlike temporal distortions, frequency-domain perturbations mainly affect the structural relationships across feature channels, rather than introducing simple point-wise shifts. Therefore, instead of enforcing strict feature matching, we aim to preserve second-order relational structures in the frequency domain.

Specifically, given the frame-level features $\bm{F}_c, \bm{F}_d \in \mathbb{R}^{T \times D}$, we first reorganize them into frequency-oriented representations:
\begin{equation}
	\bm{\tilde{F}}_c^{f} = \phi(\bm{F}_c^{f}), \quad
	\bm{\tilde{F}}_d^{f} = \phi(\bm{F}_d^{f}),
\end{equation}

Subsequently, the projected features are flattened:
\begin{equation}
	\bm{X}_c = \mathrm{Flatten}(\bm{\tilde{F}}_c^{f}), \quad
	\bm{X}_d = \mathrm{Flatten}(\bm{\tilde{F}}_d^{f}),
\end{equation}

We then adopt linear CKA to measure the similarity of second-order structures:
\begin{equation}
	\mathrm{CKA}(\bm{X}_c, \bm{X}_d)
	=
	\frac{
		\mathrm{HSIC}(\bm{K}_c, \bm{K}_d)
	}{
		\sqrt{
			\mathrm{HSIC}(\bm{K}_c, \bm{K}_c)
			\mathrm{HSIC}(\bm{K}_d, \bm{K}_d)
		}
	}.
\end{equation}

The corresponding loss is:
\begin{equation}
	\mathcal{L}_{f}
	=
	1 - \mathrm{CKA}(\bm{X}_c, \bm{X}_d).
\end{equation}

CKA provides a practical surrogate for preserving second-order frequency-domain structure, enabling the model to maintain stable spectral relationships under AFE-induced distortions.

\subsection{Training Objective}

The overall training objective combines classification and consistency constraints:
\begin{equation}
	\mathcal{L}
	=
	\mathcal{L}_{ce}^{c}
	+
	\mathcal{L}_{ce}^{d}
	+
	\lambda( \mathcal{L}_{t}
	+ \mathcal{L}_{f}).
\end{equation}

Here, $\mathcal{L}_{ce}^{c}$ and $\mathcal{L}_{ce}^{d}$ denote the cross-entropy losses for clean and distorted speech, respectively, which preserve spoof discriminability. $\mathcal{L}_{t}$ is the temporal consistency loss that reduces temporal dependency shift, while $\mathcal{L}_{f}$ is the frequency structural consistency loss that mitigates frequency structural distortion. The hyperparameter $\lambda$ balances the contribution of the consistency constraints. By jointly optimizing these objectives, the model learns spoofing representations that are both discriminative and robust to AFE-induced distortions.

\begin{table*}[t]
	\centering
	\small
	\caption{Performance comparison of different SDD models under various acoustic front-end (AFE) conditions. The settings from echo to vad correspond to intermediate stages of the AFE pipeline illustrated in Fig.~\ref{fig:afe}. \textbf{Bold} indicates the best result, and \uwave{wavy underline} indicates the second-best result.}
	\label{tab:afe_main_results}
	\renewcommand{\arraystretch}{0.92} 
	\resizebox{\textwidth}{!}{
		\begin{tabular}{c|c|cc|cc|cc|cc|cc|cc|cc}
			\specialrule{1.2pt}{0pt}{0pt}
			\multicolumn{1}{c|}{\multirow{3}{*}{\textbf{Model}}} & \multirow{3}{*}{\textbf{Pub}} & \multicolumn{2}{c|}{\multirow{2}{*}{\textbf{clean}}} & \multicolumn{12}{c}{\textbf{AFE pipeline}} \\ \cline{5-16} 
			\multicolumn{1}{c|}{} & & \multicolumn{2}{c|}{} & \multicolumn{2}{c|}{\textbf{echo}} & \multicolumn{2}{c|}{\textbf{aec}} & \multicolumn{2}{c|}{\textbf{noisy}} & \multicolumn{2}{c|}{\textbf{ns}} & \multicolumn{2}{c|}{\textbf{agc}} & \multicolumn{2}{c}{\textbf{vad}} \\ \cline{3-16} 		
			\multicolumn{1}{c|}{} & & \textbf{EER}$\downarrow$ & \textbf{AUC}$\uparrow$ & \textbf{EER}$\downarrow$ & \textbf{AUC}$\uparrow$ & \textbf{EER}$\downarrow$ & \textbf{AUC}$\uparrow$ & \textbf{EER}$\downarrow$ & \textbf{AUC}$\uparrow$ & \textbf{EER}$\downarrow$ & \textbf{AUC}$\uparrow$ & \textbf{EER}$\downarrow$ & \textbf{AUC}$\uparrow$ & \textbf{EER}$\downarrow$ & \textbf{AUC}$\uparrow$ \\ \hline
			
			AASIST \cite{jung2022aasist} & ICASSP 2022 & 0.83 &  99.93& 10.23 & 95.41 & 5.20 & 98.75 & 17.55 & 90.41 & 15.22 & 92.16 & 43.56 & 60.09 & 47.11 & 54.97 \\ \hline
			
			XLSR\_AASIST \cite{tak2022automatic} & Odyssey 2022 & 0.23 & 99.97 &  \uwave{2.20} &  \uwave{99.68} &  \uwave{2.34} &  \uwave{99.54} & \uwave{ 6.31}&  \uwave{97.84}&  \uwave{5.81} &  \uwave{97.84}&  \uwave{12.70} &  \uwave{93.69} & 22.20 & 86.59 \\ \hline
			
			XLSR\_TCM \cite{truong2024temporal} & Interspeech 2024 & 0.23 & 99.99 & 5.44 & 98.45 & 17.61 & 90.06 & 25.37 & 81.90 & 22.73 & 84.51 & 30.46 & 75.71 & 38.41 & 66.35 \\ \hline
			
			XLSR\_SLS \cite{zhang2024audio} & ACM MM 2024 & \uwave{0.23} &  \uwave{99.99} & 3.95 & 99.26 & 9.46 & 96.71 & 11.11 & 95.56 & 10.80 & 95.73 & 14.34 & 92.64 & 20.27 & 86.90 \\ \hline
			
			XLSR\_Nes2Net \cite{liu2025nes2net} & TIFS 2025 & \textbf{0.17} & \textbf{99.99} & 3.75 & 99.31 & 4.17 & 99.12 & 7.02 & 97.49 & 6.91 & 97.63 & 14.41 & 92.04 & 24.83 & 82.11 \\ \hline
			
			XLSR\_MultiConv \cite{tran2025multi} & ACM MM 2025 & 0.87 & 99.94 & 6.09 & 98.88 & 6.59 & 98.32 & 8.78 & 97.15 & 8.18 & 97.47 & 14.40 & 93.36 &  \uwave{16.83} &  \uwave{91.29} \\ \hline
			
			ALLM4ADD* \cite{gu2025allm4add} & ACM MM 2025 & 0.67 & 99.84 & 16.46 & 91.44 & 45.16 & 56.71 & 37.78 & 65.32 & 38.72 & 65.32 & 42.62& 59.81& 45.26 & 56.14 \\ \hline
			
			\textbf{Ours} &--& 0.55& 99.96 & \textbf{1.77} & \textbf{99.80}& \textbf{2.19} & \textbf{99.68} & \textbf{3.87} & \textbf{99.14} & \textbf{3.91} & \textbf{99.20} & \textbf{4.84} & \textbf{98.79} & \textbf{9.78} & \textbf{96.40}   \\ \specialrule{1.2pt}{0pt}{0pt}
		\end{tabular}
	}
	\begin{tablenotes}
		\footnotesize \item $^{*}$ denotes reproduced results; all other results are obtained using publicly available pretrained weights.
	\end{tablenotes}
\end{table*}

\section{Experimental Setup and Results Analysis}

\subsection{Dataset and acoustic front-end simulation}
\label{subsec:sim}
To thoroughly analyze the impact of environmental noise and individual components of acoustic front-end processing for SDD, we adopt the ASVspoof2019 LA \cite{todisco2019asvspoof} dataset as the research benchmark. This dataset is constructed from high-quality clean speech data. It is divided into three subsets: training, development, and evaluation. Notably, the evaluation set introduces unseen conditions in both speaker identity and spoofing algorithms.

We follow the processing pipeline illustrated in Fig.~3 to sequentially process the original speech data. First, in the echo simulation stage, for the training and development sets, we adopt a room acoustics simulation-based method\footnote{https://github.com/LCAV/pyroomacoustics}~\cite{scheibler2018pyroomacoustics}. Specifically, a virtual acoustic environment is constructed, and the multi-path propagation from the sound source to the microphone is simulated to generate speech signals with echo effects. This process can be approximately formulated as:
\begin{equation}
	x_{\text{echo}}(t) = \sum_{k} \alpha_k \, x(t - \tau_k),
\end{equation}
where $\tau_k$ and $\alpha_k$ denote the delay and attenuation associated with the $k$-th propagation path, respectively.

For the evaluation set, to more realistically simulate echo effects in practical scenarios, we adopt a room impulse response (RIR)-based\footnote{https://openslr.org/} echo simulation method. Specifically, the original speech signal is convolved with the RIR to generate the echo component, followed by introducing temporal delay and energy attenuation. The processed echo signal is then superimposed onto the original speech to form the final microphone signal. The process can be formulated as:
\begin{equation}
	x_{\text{mic}}(t) = x(t) + \beta \, (x * h)(t - \tau),
\end{equation}
where $h(t)$ denotes the room impulse response, $*$ represents convolution, $\tau$ is the echo delay, and $\beta$ is an attenuation factor controlling the echo energy. The resulting $x_{\text{mic}}(t)$ corresponds to the simulated speech signal with echo effects.

In the noise addition stage, for the training and development sets, we employ the MUSAN\footnote{https://www.openslr.org/17/} \cite{snyder2015musan} noise corpus and perform data augmentation by randomly sampling the SNR within the range of 5–20 dB. For the evaluation set, to more comprehensively assess the model’s robustness under realistic and complex acoustic conditions, we adopt the DNS\footnote{https://github.com/microsoft/DNS-Challenge} noise dataset, which consists of two noise sources: AudioSet~\cite{gemmeke2017audio} and Freesound~\cite{fonseca2017freesound}. A round-robin strategy is applied to traverse different noise samples, while the SNR is controlled to follow a uniform distribution.

Finally, we adopt the standard implementations provided by WebRTC\footnote{https://github.com/mail2chromium/Android-Audio-Processing-Using-WebRTC} to simulate the aforementioned acoustic front-end modules in a cascaded manner, thereby constructing a speech processing pipeline that closely resembles real-world communication systems. This design not only more faithfully reflects the distortion processes encountered in practical applications, but also provides a controlled and unified experimental framework for analyzing the impact of different front-end processing stages on spoofing detection performance.

\begin{figure}[!t]  % !t = 强制放在页面顶部
	\centering
	\includegraphics[scale=0.3]{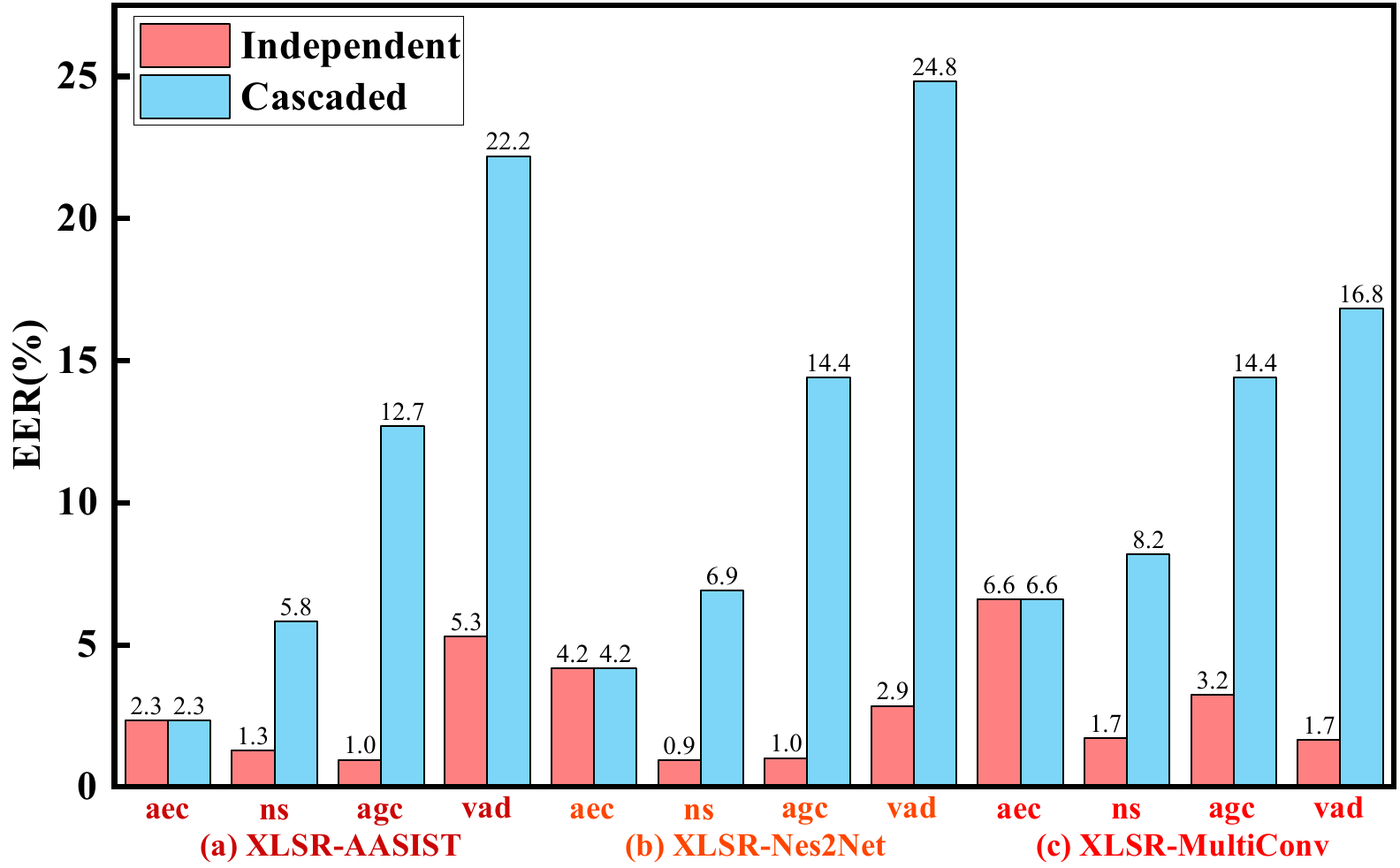}
	\caption{Performance Comparison of Different Models under Independent and Cascaded AFE Modules.}
	\label{fig:in}
\end{figure}

\subsection{Model setup}
To evaluate the performance of deepfake speech detection, we adopt three advanced models as baseline systems, namely XLSR-AASIST\footnote{https://github.com/TakHemlata/SSL\_Anti-spoofing} \cite{tak2022automatic}, XLSR-MultiConv\footnote{https://github.com/hoanmyTran/dissimilarity\_deepfake\_detection}~\cite{tran2025multi}, and XLSR-Nes2Net\footnote{https://github.com/Liu-Tianchi/Nes2Net\_ASVspoof\_ITW} \cite{liu2025nes2net}. Specifically, XLSR \cite{babu2022xls} serves as the front-end feature extractor to capture general speech representations, while AASIST, MultiConv, and Nes2Net are employed as back-end classifiers for spoofing detection. During training, RawBoost \cite{tak2022rawboost} is applied to the clean speech branch as a data augmentation strategy to enhance robustness under complex acoustic conditions.

All models are optimized using the Adam optimizer, with a learning rate of $1\times10^{-6}$ and a weight decay of $1\times10^{-4}$. Training is conducted for up to 100 epochs with an early stopping strategy, where training is terminated if no performance improvement is observed for 10 consecutive epochs. During evaluation, Equal Error Rate (EER) and Area Under the Curve (AUC) are adopted as performance metrics. Hyperparameter $\lambda$ is 0.3. All experiments are conducted on an NVIDIA RTX 4090 GPU.

\begin{table*}[ht]
	\centering
	\small
	
	\caption{Ablation study under different AFE conditions. \textbf{Bold values denote the best performance within each group.}}
	\label{tab:abl_results}
	\setlength{\tabcolsep}{3pt} 
	\renewcommand{\arraystretch}{0.92} 
	
	\begin{tabular}{c|cc|cc|cc|cc|cc|cc|cc} 
		\specialrule{1.1pt}{0pt}{0pt}
		\multirow{3}{*}{\textbf{Model}} & \multicolumn{2}{c|}{\multirow{2}{*}{\textbf{clean}}} & \multicolumn{12}{c}{\textbf{AFE pipeline}} \\ \cline{4-15} 
		& \multicolumn{2}{c|}{} & \multicolumn{2}{c|}{\textbf{echo}} & \multicolumn{2}{c|}{\textbf{aec}} & \multicolumn{2}{c|}{\textbf{noisy}} & \multicolumn{2}{c|}{\textbf{ns}} & \multicolumn{2}{c|}{\textbf{agc}} & \multicolumn{2}{c}{\textbf{vad}} \\ \cline{2-15}
		& \textbf{EER}$\downarrow$ & \textbf{AUC}$\uparrow$ & \textbf{EER}$\downarrow$ & \textbf{AUC}$\uparrow$ & \textbf{EER}$\downarrow$ & \textbf{AUC}$\uparrow$ & \textbf{EER}$\downarrow$ & \textbf{AUC}$\uparrow$ & \textbf{EER}$\downarrow$ & \textbf{AUC}$\uparrow$ & \textbf{EER}$\downarrow$ & \textbf{AUC}$\uparrow$ & \textbf{EER}$\downarrow$ & \textbf{AUC}$\uparrow$ \\ \hline
		
		% Train Data Section
		\rowcolor{base}
		\multicolumn{15}{l}{\textit{\textbf{XLSR-AASIST backbone with different training data}}} \\ \hline
		Clean-only & \textbf{0.23} &\textbf{99.97} & 2.20 & 99.68 & 2.34 & 99.54 & 6.31 & 97.84 & 5.81 & 97.84 & 12.70 & 93.69 & 22.20 & 86.59 \\ \hline
		AFE-only & 7.06 & 97.30 & 8.22 & 96.58 & 6.13 & 97.96 & 9.92 & 95.92 & 9.75 & 95.99 & 9.12 & 96.36 & 11.84 & 94.81 \\ \hline
		Mix & 0.52 & 99.95 & 2.62 & 99.46 & 5.01 & 98.99 & 6.95 & 97.93 & 6.31 & 98.31 & 7.15 & 97.83 & 17.39 & 90.33 \\ \hline
		\textbf{Ours} & 0.55 & 99.96 & \textbf{1.77} & \textbf{99.80} & \textbf{2.19} & \textbf{99.68} & \textbf{3.87} & \textbf{99.14} & \textbf{3.91} & \textbf{99.20 }& \textbf{4.84 }& \textbf{98.79 }& \textbf{9.78} & \textbf{96.40} \\ \hline
		
		% Backbone Section
		\rowcolor{cmp}
		\multicolumn{15}{l}{\textit{\textbf{Other backbones trained on mixed data}}} \\ \hline
		XLSR\_Nes2Net \textit{\scriptsize w/ TFCL} &  0.56&  99.97 &  \textbf{1.89} &  \textbf{99.78} &  \textbf{2.44} &  \textbf{99.66} &  \textbf{4.61} &  \textbf{98.98} &  \textbf{5.04} &  \textbf{98.76} &  \textbf{5.91} &  \textbf{98.32} &  \textbf{11.08} &  \textbf{95.59} \\ \hline
		XLSR\_Nes2Net \textit{\scriptsize w/o TFCL} & \textbf{0.46} & \textbf{99.98} & 4.35 & 99.12 & 11.98 & 95.26 & 15.82 & 92.13 & 14.49 & 93.26 & 12.17 & 94.92 & 15.58 & 92.32 \\ \hline
		XLSR\_MultiConv \textit{\scriptsize w/ TFCL} & 0.76 & 99.86 & 2.80 & 99.50 & 3.49 & 99.27 & 5.95 & 98.33 & 5.44 & 98.56 & 7.12 & 97.88 & 11.76 & 94.61 \\ \hline
		XLSR\_MultiConv \textit{\scriptsize w/o TFCL} & 0.91 & 99.90 & 4.30 & 99.12 & 5.94 & 98.64 & 8.73 & 97.42 & 8.49 & 97.43 & 9.37 & 96.85 & 14.76 & 92.82 \\ \hline
		
		% Ablation Section
		\rowcolor{abl}
		\multicolumn{15}{l}{\textit{\textbf{Ablation experiments (XLSR-AASIST)}}} \\ \hline
		TFCL & 0.55 & 99.96 & 1.77 & 99.80 & 2.19 & 99.68 & \textbf{3.87} &\textbf{ 99.14} & \textbf{3.91} & \textbf{99.20}& \textbf{4.84} & \textbf{98.79} & \textbf{9.78} & \textbf{96.40} \\ \hline
		w/o TACL &0.71 & 99.92& 2.12&99.59 &3.10 &99.40 &5.29 &98.44 & 5.15& 98.64&5.78 &98.24 &10.48 &95.31 \\ \hline
		w/o Bi\_attention &0.58&\textbf{99.97} &2.19 &99.65& 2.80&99.53 &7.08 &97.66 &5.22 &98.62 &5.91 &98.23 &11.37 &94.91   \\ \hline
		w/o Attention &\textbf{0.45} &99.96 &\textbf{1.59} &\textbf{99.80} &\textbf{2.05} &\textbf{99.70} &4.56 &98.94 &5.03 &98.90 &5.85 &98.52 &11.56 &95.39 \\ \hline
		w/o FSCL &0.72 & 99.92& 2.58& 99.63& 3.26& 99.49&5.33 &98.66 &5.18 &98.73 &6.15 &98.30 & 10.99& 96.16\\ 
		
		\specialrule{1.2pt}{0pt}{0pt}
	\end{tabular}
	
\end{table*}

\subsection{Robustness of state-of-the-art models under AFE-Induced degradations}
Table~\ref{tab:afe_main_results} presents a performance comparison of various state-of-the-art (SOTA) SDD models under different AFE processing stages. The evaluation covers scenarios ranging from the clean condition to intermediate stages of the AFE pipeline (corresponding to the stages from Echo to VAD in Fig.~\ref{fig:afe}), enabling a systematic analysis of model robustness under real-world communication processing pipelines.

\textbf{Significant performance gap: robustness degradation from ideal to real-world conditions.}
The experimental results show that although existing models achieve excellent performance under the clean condition (with very low EER and near-saturated AUC), their performance degrades significantly and consistently once AFE processing is introduced. As the processing progresses from Echo to VAD, this degradation exhibits a clear accumulative trend. For instance, the EER of AASIST increases from 0.83\% to 47.11\%, while that of the SSL-based XLSR\_AASIST rises from 0.23\% to 22.20\%. These observations indicate that the high performance achieved under ideal conditions does not directly transfer to real-world communication scenarios, and AFE processing has become a key factor limiting the practical deployment of SDD models.

\textbf{Cascaded distortion effect: cumulative degradation of discriminative features under AFE processing.}  
Further analysis reveals that the performance degradation is not caused by individual processing modules independently, but rather results from the cascaded effect of the AFE pipeline. The stage-wise results in Table~\ref{tab:afe_main_results}, from AEC and NS to AGC and VAD, already exhibit a clear accumulative degradation trend.
To further investigate the source of this cascaded effect, Fig.~\ref{fig:in} compares model performance under independent AFE modules and cascaded processing conditions. It can be observed that when individual AFE modules are applied independently, the performance degradation remains relatively limited. In contrast, under cascaded processing, the degradation is significantly amplified, far exceeding the simple sum of individual effects. For example, under AGC and VAD conditions, the EER increase caused by cascaded processing is substantially larger than that induced by the corresponding independent modules, indicating strong coupling effects across processing stages.
These observations suggest that AFE processing not only introduces local perturbations, but also progressively amplifies and accumulates distortions through cross-module interactions. For instance, AEC may introduce residual artifacts or information loss, NS alters local spectral structures, AGC redistributes energy and amplifies existing distortions, while VAD further modifies temporal structures. Under cascaded conditions, these effects interact and compound, ultimately disrupting the subtle discriminative cues in spoofed speech, making it difficult for methods relying on single-feature patterns or local structures to maintain stable performance.

\textbf{Robustness improvement: stability of the proposed method under AFE conditions.}  
In contrast, the proposed method exhibits more stable performance across all AFE conditions. It effectively mitigates performance degradation both in the early stages of the processing pipeline (Echo, AEC) and in more challenging later stages (AGC, VAD), maintaining lower EER and higher AUC. This suggests that, by jointly modeling temporal dependency consistency and frequency structural consistency, the model is able to learn robust spoof-discriminative representations that remain invariant before and after AFE processing, thereby improving stability under cascaded distortions.

\begin{figure*}[!t]  % !t = 强制放在页面顶部
	\centering
	\includegraphics[scale=0.37]{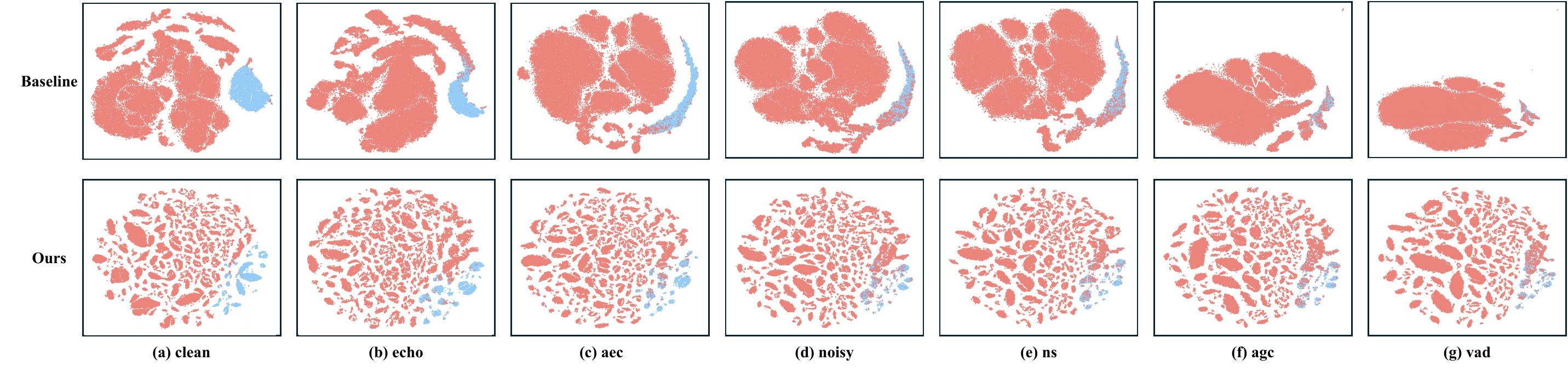}
	\caption{t-SNE~\cite{van2008visualizing} visualization of feature distributions across different AFE stages, with XLSR+AASIST as the baseline model.}
	\label{fig:sne}
\end{figure*}

\begin{figure}[!t]  % !t = 强制放在页面顶部
	\centering
	\includegraphics[scale=0.42]{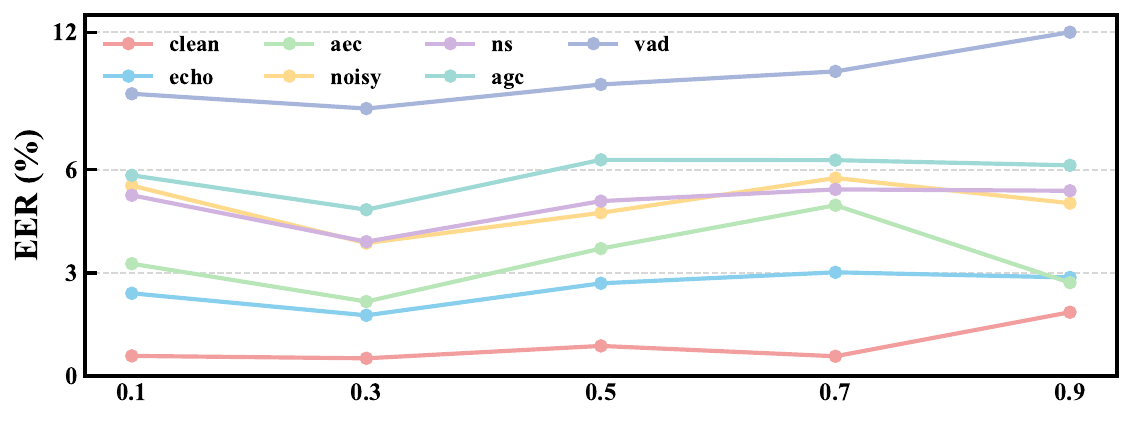}
	\caption{Results of hyperparameter $\lambda$ under various AFE conditions.}
	\label{fig:hyp}
\end{figure}

\subsection{Ablation study analysis}

Table~\ref{tab:abl_results} presents the ablation results under different AFE conditions, covering various training data strategies, model frameworks, and component-level designs. The evaluation spans from the clean condition to different stages of the AFE pipeline, providing a comprehensive analysis of how each factor contributes to robustness. Based on these results, several key observations can be drawn:

\textbf{(1) Impact of training data strategies.}  
Under the XLSR-AASIST framework, different training data strategies lead to significantly different behaviors. The model trained with clean-only data achieves the best performance under the clean condition, but degrades severely under AFE processing. The AFE-only strategy improves performance under certain AFE conditions but remains unstable overall. The Mix strategy partially alleviates the degradation, yet still suffers noticeable performance drops under more severe distortions (e.g., AGC and VAD). In contrast, the proposed method achieves consistently stable performance across all AFE conditions, indicating that simple data augmentation or mixing is insufficient to address the distribution shift introduced by AFE processing.

\textbf{(2) Effectiveness across different frameworks.}  
The proposed TFCL framework consistently improves performance across different model frameworks, such as Nes2Net and MultiConv. Compared to their corresponding w/o TFCL counterparts, models equipped with TFCL achieve lower EER under all AFE conditions. Notably, the improvement becomes more pronounced as distortions accumulate along the AFE pipeline. This demonstrates that the proposed method is not tied to a specific architecture, but generalizes well across different frameworks.

\textbf{(3) Contribution of individual components.}  
Further ablation results highlight the roles of different components. Removing temporal consistency learning (w/o TACL) leads to a clear performance drop. Additionally, removing the attention mechanisms within TACL (w/o Bi-attention and w/o Attention) also results in noticeable degradation, indicating that attention-based soft alignment is essential for cross-condition feature matching. On top of this, removing frequency structural consistency learning (w/o FSCL) further degrades performance, demonstrating that frequency-domain constraints provide complementary benefits. Overall, TACL and FSCL operate on temporal dependency and frequency structure, respectively, enabling the model to maintain stable representations under AFE-induced distortions.

\subsection{t-SNE visualization analysis}

Figs.~\ref{fig:sne} presents the t-SNE visualization of feature distributions for the baseline model (XLSR-AASIST) and the proposed method under different AFE processing stages. The figure intuitively illustrates how model representations evolve in the feature space as the AFE pipeline progresses from echo to vad.

For the baseline model, under the clean condition, the feature distributions exhibit relatively compact cluster structures, indicating that the model is able to learn clear decision boundaries in ideal scenarios. However, as AFE processing is introduced and progressively intensified, these structures rapidly deform and undergo noticeable distribution shifts. The feature clusters become increasingly stretched and even overlap, leading to a significant degradation in class separability. This suggests that the baseline model tends to fit decision boundaries under specific data distributions, and the learned discriminative patterns are highly sensitive to perturbations introduced by AFE processing. Once the input distribution changes, the model performance becomes difficult to maintain.

In contrast, the proposed method exhibits more stable feature distribution structures across different AFE conditions. Although the feature distributions are relatively more dispersed under the clean condition, their overall structure changes less as the processing pipeline progresses, and the relative arrangement between different classes is better preserved. This observation indicates that the proposed method is more inclined to learn representation spaces that remain invariant across conditions, rather than relying on decision boundaries tailored to specific input distributions.

In summary, the proposed time-frequency consistency learning framework effectively mitigates the distribution shift induced by AFE processing, enabling the model to maintain stable discriminative capability under cascaded distortions and thereby improving robustness in real-world communication scenarios.

\subsection{Hyperparameter experimental analysis}
Fig.~\ref{fig:hyp} illustrates the performance of different consistency weights $\lambda$ under various AFE conditions. Overall, the model performance is clearly influenced by the choice of $\lambda$. When $\lambda$ is small, the consistency constraint is insufficient, and the model tends to rely more on the classification objective, making it less effective in mitigating the distribution shift introduced by AFE processing. In contrast, when $\lambda$ is too large, the overly strong constraint restricts the discriminative capability of the model, leading to performance degradation. Across different AFE conditions, $\lambda = 0.3$ achieves the most stable performance, maintaining low EER while balancing robustness across all processing stages.

\section{Conclusion}

In this work, we revisit the robustness problem of SDD from a more realistic deployment perspective. While prior studies primarily focus on additive noise and communication-related degradations, they largely overlook the impact of AFE processing pipelines that are ubiquitous in modern RTC systems. To bridge this gap, we construct a unified AFE simulation pipeline and systematically evaluate existing SDD models across different processing stages. Our analysis reveals that AFE-induced distortions exhibit strong time-frequency coupling characteristics and are progressively amplified through cascaded processing, leading to severe degradation in spoof-discriminative representations.

To address these challenges, we propose a TFCL framework that explicitly models the invariance of spoof-discriminative representations before and after AFE processing. Specifically, the proposed method mitigates temporal dependency disruption via attention-based soft alignment and alleviates frequency structural distortion through consistency constraints in the spectral domain. By jointly enforcing temporal and frequency consistency, the model learns more robust representations that generalize across AFE-induced distribution shifts. Extensive experiments demonstrate that the proposed approach consistently improves robustness under various AFE conditions and across different model frameworks.

Despite these improvements, this work represents an initial step toward understanding SDD under realistic communication pipelines. In this paper, we focus on modeling the AFE processing as a unified module and systematically analyze its impact in isolation. However, in practical RTC systems, AFE processing is further coupled with network transmission effects, such as codec compression, packet loss, and bandwidth constraints, which may introduce more complex and intertwined distortions. Modeling such joint effects remains a challenging and important direction for future research.

\begin{acks}
This work is supported by the Natural Science Foundation of China (NSFC) under the grant NO.62572358,62372334.

\end{acks}

\bibliographystyle{ACM-Reference-Format}
%\balance
\bibliography{sample-base}


\begin{thebibliography}{35}

%%% the Web address.

\ifx \showCODEN    \undefined \def \showCODEN     #1{\unskip}     \fi
\ifx \showISBNx    \undefined \def \showISBNx     #1{\unskip}     \fi
\ifx \showISBNxiii \undefined \def \showISBNxiii  #1{\unskip}     \fi
\ifx \showISSN     \undefined \def \showISSN      #1{\unskip}     \fi
\ifx \showLCCN     \undefined \def \showLCCN      #1{\unskip}     \fi
\ifx \shownote     \undefined \def \shownote      #1{#1}          \fi
\ifx \showarticletitle \undefined \def \showarticletitle #1{#1}   \fi
\ifx \showURL      \undefined \def \showURL       {\relax}        \fi
% invisible to TeX
\providecommand\bibfield[2]{#2}
\providecommand\bibinfo[2]{#2}
\providecommand\natexlab[1]{#1}
\providecommand\showeprint[2][]{arXiv:#2}

\bibitem[Babu et~al\mbox{.}(2022)]%
        {babu2022xls}
\bibfield{author}{\bibinfo{person}{Arun Babu}, \bibinfo{person}{Changhan Wang},
  \bibinfo{person}{Andros Tjandra}, \bibinfo{person}{Kushal Lakhotia},
  \bibinfo{person}{Qiantong Xu}, \bibinfo{person}{Naman Goyal},
  \bibinfo{person}{Kritika Singh}, \bibinfo{person}{Patrick von Platen},
  \bibinfo{person}{Yatharth Saraf}, \bibinfo{person}{Juan Pino},
  {et~al\mbox{.}}} \bibinfo{year}{2022}\natexlab{}.
\newblock \showarticletitle{XLS-R: Self-supervised Cross-lingual Speech
  Representation Learning at Scale}. In \bibinfo{booktitle}{\emph{Proc.
  Interspeech 2022}}. \bibinfo{pages}{2278--2282}.
\newblock


\bibitem[Baevski et~al\mbox{.}(2020)]%
        {baevski2020wav2vec}
\bibfield{author}{\bibinfo{person}{Alexei Baevski}, \bibinfo{person}{Yuhao
  Zhou}, \bibinfo{person}{Abdelrahman Mohamed}, {and} \bibinfo{person}{Michael
  Auli}.} \bibinfo{year}{2020}\natexlab{}.
\newblock \showarticletitle{wav2vec 2.0: A framework for self-supervised
  learning of speech representations}.
\newblock \bibinfo{journal}{\emph{Advances in neural information processing
  systems}}  \bibinfo{volume}{33} (\bibinfo{year}{2020}),
  \bibinfo{pages}{12449--12460}.
\newblock


\bibitem[Bashir({[n.\,d.]})]%
        {webrtc_pipeline}
\bibfield{author}{\bibinfo{person}{Muhammad~Usman Bashir}.}
  \bibinfo{year}{[n.\,d.]}\natexlab{}.
\newblock \bibinfo{booktitle}{\emph{WebRTC Audio Processing Overview}}.
\newblock
\urldef\tempurl%
\url{https://github.com/mail2chromium/Android-Audio-Processing-Using-WebRTC}
\showURL{%
\tempurl}


\bibitem[Cao et~al\mbox{.}(2025)]%
        {cao2025robust}
\bibfield{author}{\bibinfo{person}{Jialu Cao}, \bibinfo{person}{Hui Tian},
  \bibinfo{person}{Peng Tian}, \bibinfo{person}{Haizhou Li}, {and}
  \bibinfo{person}{Jianzong Wang}.} \bibinfo{year}{2025}\natexlab{}.
\newblock \showarticletitle{Robust Detection of Partially Spoofed Audio Using
  Semantic-Aware Inconsistency Learning}.
\newblock \bibinfo{journal}{\emph{IEEE Transactions on Audio, Speech and
  Language Processing}} (\bibinfo{year}{2025}).
\newblock


\bibitem[{Channel News Asia}(2025)]%
        {cna_deepfake_scam_2025}
\bibfield{author}{\bibinfo{person}{{Channel News Asia}}.}
  \bibinfo{year}{2025}\natexlab{}.
\newblock \bibinfo{booktitle}{\emph{Company Finance Director Nearly Loses Over
  {US}\$499{,}000 to Scammers Using Deepfake to Impersonate CEO}}.
\newblock
\urldef\tempurl%
\url{https://www.channelnewsasia.com/singapore/deepfake-scam-impersonate-ceo-company-finance-director-5048706}
\showURL{%
\tempurl}
\newblock
\shownote{Accessed: 2025-11-6}.


\bibitem[Chen et~al\mbox{.}(2025b)]%
        {chen2025adaptive}
\bibfield{author}{\bibinfo{person}{Qixian Chen}, \bibinfo{person}{Yuxiong Xu},
  \bibinfo{person}{Sara Mandelli}, \bibinfo{person}{Sheng Li}, {and}
  \bibinfo{person}{Bin Li}.} \bibinfo{year}{2025}\natexlab{b}.
\newblock \showarticletitle{Adaptive Mixture of Low-Rank Experts for Robust
  Audio Spoofing Detection}.
\newblock \bibinfo{journal}{\emph{IEEE Signal Processing Letters}}
  (\bibinfo{year}{2025}).
\newblock


\bibitem[Chen et~al\mbox{.}(2025a)]%
        {chen25j_interspeech}
\bibfield{author}{\bibinfo{person}{Xuanjun Chen}, \bibinfo{person}{I-Ming Lin},
  \bibinfo{person}{Lin Zhang}, \bibinfo{person}{Jiawei Du},
  \bibinfo{person}{Haibin Wu}, \bibinfo{person}{Hung yi Lee}, {and}
  \bibinfo{person}{Jyh-Shing~Roger Jang}.} \bibinfo{year}{2025}\natexlab{a}.
\newblock \showarticletitle{{Codec-Based Deepfake Source Tracing via Neural
  Audio Codec Taxonomy}}. In \bibinfo{booktitle}{\emph{{Interspeech 2025}}}.
  \bibinfo{pages}{1538--1542}.
\newblock
\showISSN{2958-1796}
\href{https://doi.org/10.21437/Interspeech.2025-1297}{doi:\nolinkurl{10.21437/Interspeech.2025-1297}}


\bibitem[Chen et~al\mbox{.}(2024)]%
        {chen2024rawbmamba}
\bibfield{author}{\bibinfo{person}{Yujie Chen}, \bibinfo{person}{Jiangyan Yi},
  \bibinfo{person}{Jun Xue}, \bibinfo{person}{Chenglong Wang},
  \bibinfo{person}{Xiaohui Zhang}, \bibinfo{person}{Shunbo Dong},
  \bibinfo{person}{Siding Zeng}, \bibinfo{person}{Jianhua Tao},
  \bibinfo{person}{Zhao Lv}, {and} \bibinfo{person}{Cunhang Fan}.}
  \bibinfo{year}{2024}\natexlab{}.
\newblock \showarticletitle{RawBMamba: End-to-End Bidirectional State Space
  Model for Audio Deepfake Detection}. In \bibinfo{booktitle}{\emph{Proc.
  Interspeech 2024}}. \bibinfo{pages}{2720--2724}.
\newblock


\bibitem[Du et~al\mbox{.}(2024)]%
        {du2024cosyvoice}
\bibfield{author}{\bibinfo{person}{Zhihao Du}, \bibinfo{person}{Qian Chen},
  \bibinfo{person}{Shiliang Zhang}, \bibinfo{person}{Kai Hu},
  \bibinfo{person}{Heng Lu}, \bibinfo{person}{Yexin Yang},
  \bibinfo{person}{Hangrui Hu}, \bibinfo{person}{Siqi Zheng},
  \bibinfo{person}{Yue Gu}, \bibinfo{person}{Ziyang Ma}, {et~al\mbox{.}}}
  \bibinfo{year}{2024}\natexlab{}.
\newblock \showarticletitle{Cosyvoice: A scalable multilingual zero-shot
  text-to-speech synthesizer based on supervised semantic tokens}.
\newblock \bibinfo{journal}{\emph{arXiv preprint arXiv:2407.05407}}
  (\bibinfo{year}{2024}).
\newblock


\bibitem[Fan et~al\mbox{.}(2024)]%
        {fan2024dual}
\bibfield{author}{\bibinfo{person}{Cunhang Fan}, \bibinfo{person}{Mingming
  Ding}, \bibinfo{person}{Jianhua Tao}, \bibinfo{person}{Ruibo Fu},
  \bibinfo{person}{Jiangyan Yi}, \bibinfo{person}{Zhengqi Wen}, {and}
  \bibinfo{person}{Zhao Lv}.} \bibinfo{year}{2024}\natexlab{}.
\newblock \showarticletitle{Dual-branch knowledge distillation for noise-robust
  synthetic speech detection}.
\newblock \bibinfo{journal}{\emph{IEEE/ACM Transactions on Audio, Speech, and
  Language Processing}}  \bibinfo{volume}{32} (\bibinfo{year}{2024}),
  \bibinfo{pages}{2453--2466}.
\newblock


\bibitem[Fonseca et~al\mbox{.}(2017)]%
        {fonseca2017freesound}
\bibfield{author}{\bibinfo{person}{Eduardo Fonseca}, \bibinfo{person}{Jordi
  Pons~Puig}, \bibinfo{person}{Xavier Favory}, \bibinfo{person}{Frederic
  Font~Corbera}, \bibinfo{person}{Dmitry Bogdanov}, \bibinfo{person}{Andres
  Ferraro}, \bibinfo{person}{Sergio Oramas}, \bibinfo{person}{Alastair Porter},
  {and} \bibinfo{person}{Xavier Serra}.} \bibinfo{year}{2017}\natexlab{}.
\newblock \showarticletitle{Freesound datasets: a platform for the creation of
  open audio datasets}. In \bibinfo{booktitle}{\emph{Hu X, Cunningham SJ,
  Turnbull D, Duan Z, editors. Proceedings of the 18th ISMIR Conference; 2017
  oct 23-27; Suzhou, China.[Canada]: International Society for Music
  Information Retrieval; 2017. p. 486-93.}} International Society for Music
  Information Retrieval (ISMIR).
\newblock


\bibitem[Gemmeke et~al\mbox{.}(2017)]%
        {gemmeke2017audio}
\bibfield{author}{\bibinfo{person}{Jort~F Gemmeke}, \bibinfo{person}{Daniel~PW
  Ellis}, \bibinfo{person}{Dylan Freedman}, \bibinfo{person}{Aren Jansen},
  \bibinfo{person}{Wade Lawrence}, \bibinfo{person}{R~Channing Moore},
  \bibinfo{person}{Manoj Plakal}, {and} \bibinfo{person}{Marvin Ritter}.}
  \bibinfo{year}{2017}\natexlab{}.
\newblock \showarticletitle{Audio set: An ontology and human-labeled dataset
  for audio events}. In \bibinfo{booktitle}{\emph{2017 IEEE international
  conference on acoustics, speech and signal processing (ICASSP)}}. IEEE,
  \bibinfo{pages}{776--780}.
\newblock


\bibitem[Google({[n.\,d.]})]%
        {webrtc_apm}
\bibfield{author}{\bibinfo{person}{Google}.}
  \bibinfo{year}{[n.\,d.]}\natexlab{}.
\newblock \bibinfo{booktitle}{\emph{WebRTC Audio Processing Module}}.
\newblock
\urldef\tempurl%
\url{https://webrtc.googlesource.com/src/+/refs/heads/main/modules/audio_processing/g3doc/audio_processing_module.md}
\showURL{%
\tempurl}


\bibitem[Gu et~al\mbox{.}(2025)]%
        {gu2025allm4add}
\bibfield{author}{\bibinfo{person}{Hao Gu}, \bibinfo{person}{Jiangyan Yi},
  \bibinfo{person}{Chenglong Wang}, \bibinfo{person}{Jianhua Tao},
  \bibinfo{person}{Zheng Lian}, \bibinfo{person}{Jiayi He},
  \bibinfo{person}{Yong Ren}, \bibinfo{person}{Yujie Chen}, {and}
  \bibinfo{person}{Zhengqi Wen}.} \bibinfo{year}{2025}\natexlab{}.
\newblock \showarticletitle{Allm4add: Unlocking the capabilities of audio large
  language models for audio deepfake detection}. In
  \bibinfo{booktitle}{\emph{Proceedings of the 33rd ACM International
  Conference on Multimedia}}. \bibinfo{pages}{11736--11745}.
\newblock


\bibitem[Jung et~al\mbox{.}(2022)]%
        {jung2022aasist}
\bibfield{author}{\bibinfo{person}{Jee-weon Jung}, \bibinfo{person}{Hee-Soo
  Heo}, \bibinfo{person}{Hemlata Tak}, \bibinfo{person}{Hye-jin Shim},
  \bibinfo{person}{Joon~Son Chung}, \bibinfo{person}{Bong-Jin Lee},
  \bibinfo{person}{Ha-Jin Yu}, {and} \bibinfo{person}{Nicholas Evans}.}
  \bibinfo{year}{2022}\natexlab{}.
\newblock \showarticletitle{Aasist: Audio anti-spoofing using integrated
  spectro-temporal graph attention networks}. In
  \bibinfo{booktitle}{\emph{ICASSP 2022-2022 IEEE international conference on
  acoustics, speech and signal processing (ICASSP)}}. IEEE,
  \bibinfo{pages}{6367--6371}.
\newblock


\bibitem[Liu et~al\mbox{.}(2025)]%
        {liu2025nes2net}
\bibfield{author}{\bibinfo{person}{Tianchi Liu}, \bibinfo{person}{Duc-Tuan
  Truong}, \bibinfo{person}{Rohan~Kumar Das}, \bibinfo{person}{Kong~Aik Lee},
  {and} \bibinfo{person}{Haizhou Li}.} \bibinfo{year}{2025}\natexlab{}.
\newblock \showarticletitle{Nes2net: A lightweight nested architecture for
  foundation model driven speech anti-spoofing}.
\newblock \bibinfo{journal}{\emph{IEEE Transactions on Information Forensics
  and Security}}  \bibinfo{volume}{20} (\bibinfo{year}{2025}),
  \bibinfo{pages}{12005--12018}.
\newblock


\bibitem[Liu et~al\mbox{.}(2023)]%
        {liu2023asvspoof}
\bibfield{author}{\bibinfo{person}{Xuechen Liu}, \bibinfo{person}{Xin Wang},
  \bibinfo{person}{Md Sahidullah}, \bibinfo{person}{Jose Patino},
  \bibinfo{person}{H{\'e}ctor Delgado}, \bibinfo{person}{Tomi Kinnunen},
  \bibinfo{person}{Massimiliano Todisco}, \bibinfo{person}{Junichi Yamagishi},
  \bibinfo{person}{Nicholas Evans}, \bibinfo{person}{Andreas Nautsch},
  {et~al\mbox{.}}} \bibinfo{year}{2023}\natexlab{}.
\newblock \showarticletitle{Asvspoof 2021: Towards spoofed and deepfake speech
  detection in the wild}.
\newblock \bibinfo{journal}{\emph{IEEE/ACM Transactions on Audio, Speech, and
  Language Processing}}  \bibinfo{volume}{31} (\bibinfo{year}{2023}),
  \bibinfo{pages}{2507--2522}.
\newblock


\bibitem[Scheibler et~al\mbox{.}(2018)]%
        {scheibler2018pyroomacoustics}
\bibfield{author}{\bibinfo{person}{Robin Scheibler}, \bibinfo{person}{Eric
  Bezzam}, {and} \bibinfo{person}{Ivan Dokmani{\'c}}.}
  \bibinfo{year}{2018}\natexlab{}.
\newblock \showarticletitle{Pyroomacoustics: A python package for audio room
  simulation and array processing algorithms}. In
  \bibinfo{booktitle}{\emph{2018 IEEE international conference on acoustics,
  speech and signal processing (ICASSP)}}. IEEE, \bibinfo{pages}{351--355}.
\newblock


\bibitem[Sen et~al\mbox{.}(2025)]%
        {sen2025toward}
\bibfield{author}{\bibinfo{person}{Udayon Sen}, \bibinfo{person}{Alka Luqman},
  {and} \bibinfo{person}{Anupam Chattopadhyay}.}
  \bibinfo{year}{2025}\natexlab{}.
\newblock \showarticletitle{Toward Noise-Aware Audio Deepfake Detection:
  Survey, SNR-Benchmarks, and Practical Recipes}.
\newblock \bibinfo{journal}{\emph{arXiv preprint arXiv:2512.13744}}
  (\bibinfo{year}{2025}).
\newblock


\bibitem[Snyder et~al\mbox{.}(2015)]%
        {snyder2015musan}
\bibfield{author}{\bibinfo{person}{David Snyder}, \bibinfo{person}{Guoguo
  Chen}, {and} \bibinfo{person}{Daniel Povey}.}
  \bibinfo{year}{2015}\natexlab{}.
\newblock \showarticletitle{Musan: A music, speech, and noise corpus}.
\newblock \bibinfo{journal}{\emph{arXiv preprint arXiv:1510.08484}}
  (\bibinfo{year}{2015}).
\newblock


\bibitem[Tak et~al\mbox{.}(2022a)]%
        {tak2022rawboost}
\bibfield{author}{\bibinfo{person}{Hemlata Tak}, \bibinfo{person}{Madhu
  Kamble}, \bibinfo{person}{Jose Patino}, \bibinfo{person}{Massimiliano
  Todisco}, {and} \bibinfo{person}{Nicholas Evans}.}
  \bibinfo{year}{2022}\natexlab{a}.
\newblock \showarticletitle{Rawboost: A raw data boosting and augmentation
  method applied to automatic speaker verification anti-spoofing}. In
  \bibinfo{booktitle}{\emph{ICASSP 2022-2022 IEEE International Conference on
  Acoustics, Speech and Signal Processing (ICASSP)}}. IEEE,
  \bibinfo{pages}{6382--6386}.
\newblock


\bibitem[Tak et~al\mbox{.}(2022b)]%
        {tak2022automatic}
\bibfield{author}{\bibinfo{person}{Hemlata Tak}, \bibinfo{person}{Massimiliano
  Todisco}, \bibinfo{person}{Xin Wang}, \bibinfo{person}{Jee-weon Jung},
  \bibinfo{person}{Junichi Yamagishi}, {and} \bibinfo{person}{Nicholas Evans}.}
  \bibinfo{year}{2022}\natexlab{b}.
\newblock \showarticletitle{Automatic Speaker Verification Spoofing and
  Deepfake Detection Using Wav2vec 2.0 and Data Augmentation}. In
  \bibinfo{booktitle}{\emph{Proc. Odyssey 2022}}. \bibinfo{pages}{112--119}.
\newblock


\bibitem[Todisco et~al\mbox{.}(2019)]%
        {todisco2019asvspoof}
\bibfield{author}{\bibinfo{person}{Massimiliano Todisco}, \bibinfo{person}{Xin
  Wang}, \bibinfo{person}{Ville Vestman}, \bibinfo{person}{Md Sahidullah},
  \bibinfo{person}{Hector Delgado}, \bibinfo{person}{Andreas Nautsch},
  \bibinfo{person}{Junichi Yamagishi}, \bibinfo{person}{Nicholas Evans},
  \bibinfo{person}{Tomi Kinnunen}, {and} \bibinfo{person}{Kong~Aik Lee}.}
  \bibinfo{year}{2019}\natexlab{}.
\newblock \showarticletitle{ASVspoof 2019: Future Horizons in Spoofed and Fake
  Audio Detection}. In \bibinfo{booktitle}{\emph{Interspeech 2019}}.
  International Speech Communication Association, \bibinfo{pages}{1008--1012}.
\newblock


\bibitem[Tran et~al\mbox{.}(2025)]%
        {tran2025multi}
\bibfield{author}{\bibinfo{person}{Hoan~My Tran}, \bibinfo{person}{Damien
  Lolive}, \bibinfo{person}{Aghilas Sini}, \bibinfo{person}{Arnaud Delhay},
  \bibinfo{person}{Pierre-Fran{\c{c}}ois Marteau}, {and} \bibinfo{person}{David
  Guennec}.} \bibinfo{year}{2025}\natexlab{}.
\newblock \showarticletitle{Multi-level SSL feature gating for audio deepfake
  detection}. In \bibinfo{booktitle}{\emph{Proceedings of the 33rd ACM
  International Conference on Multimedia}}. \bibinfo{pages}{11766--11775}.
\newblock


\bibitem[Truong et~al\mbox{.}(2024)]%
        {truong2024temporal}
\bibfield{author}{\bibinfo{person}{Duc-Tuan Truong}, \bibinfo{person}{Ruijie
  Tao}, \bibinfo{person}{Tuan Nguyen}, \bibinfo{person}{Hieu-Thi Luong},
  \bibinfo{person}{Kong~Aik Lee}, {and} \bibinfo{person}{Eng~Siong Chng}.}
  \bibinfo{year}{2024}\natexlab{}.
\newblock \showarticletitle{Temporal-Channel Modeling in Multi-head
  Self-Attention for Synthetic Speech Detection}. In
  \bibinfo{booktitle}{\emph{Proc. Interspeech 2024}}.
  \bibinfo{pages}{537--541}.
\newblock


\bibitem[Van~der Maaten and Hinton(2008)]%
        {van2008visualizing}
\bibfield{author}{\bibinfo{person}{Laurens Van~der Maaten} {and}
  \bibinfo{person}{Geoffrey Hinton}.} \bibinfo{year}{2008}\natexlab{}.
\newblock \showarticletitle{Visualizing data using t-SNE.}
\newblock \bibinfo{journal}{\emph{Journal of machine learning research}}
  \bibinfo{volume}{9}, \bibinfo{number}{11} (\bibinfo{year}{2008}).
\newblock


\bibitem[Wang et~al\mbox{.}(2024)]%
        {wang2024asvspoof}
\bibfield{author}{\bibinfo{person}{Xin Wang}, \bibinfo{person}{H{\'e}ctor
  Delgado}, \bibinfo{person}{Hemlata Tak}, \bibinfo{person}{Jee-Weon Jung},
  \bibinfo{person}{Hye-Jin Shim}, \bibinfo{person}{Massimiliano Todisco},
  \bibinfo{person}{Ivan Kukanov}, \bibinfo{person}{Xuechen Liu},
  \bibinfo{person}{Md Sahidullah}, \bibinfo{person}{Tomi Kinnunen},
  {et~al\mbox{.}}} \bibinfo{year}{2024}\natexlab{}.
\newblock \showarticletitle{ASVspoof 5: crowdsourced speech data, deepfakes,
  and adversarial attacks at scale}. In \bibinfo{booktitle}{\emph{The Automatic
  Speaker Verification Spoofing Countermeasures Workshop (ASVspoof 2024)}}.
  ISCA, \bibinfo{pages}{1--8}.
\newblock


\bibitem[Wang et~al\mbox{.}(2022)]%
        {wang2022nn3a}
\bibfield{author}{\bibinfo{person}{Ziteng Wang}, \bibinfo{person}{Yueyue Na},
  \bibinfo{person}{Biao Tian}, {and} \bibinfo{person}{Qiang Fu}.}
  \bibinfo{year}{2022}\natexlab{}.
\newblock \showarticletitle{NN3A: Neural network supported acoustic echo
  cancellation, noise suppression and automatic gain control for real-time
  communications}. In \bibinfo{booktitle}{\emph{ICASSP 2022-2022 IEEE
  International Conference on Acoustics, Speech and Signal Processing
  (ICASSP)}}. IEEE, \bibinfo{pages}{661--665}.
\newblock


\bibitem[Xie et~al\mbox{.}(2025)]%
        {xie2025codecfake}
\bibfield{author}{\bibinfo{person}{Yuankun Xie}, \bibinfo{person}{Yi Lu},
  \bibinfo{person}{Ruibo Fu}, \bibinfo{person}{Zhengqi Wen},
  \bibinfo{person}{Zhiyong Wang}, \bibinfo{person}{Jianhua Tao},
  \bibinfo{person}{Xin Qi}, \bibinfo{person}{Xiaopeng Wang},
  \bibinfo{person}{Yukun Liu}, \bibinfo{person}{Haonan Cheng}, {et~al\mbox{.}}}
  \bibinfo{year}{2025}\natexlab{}.
\newblock \showarticletitle{The codecfake dataset and countermeasures for the
  universally detection of deepfake audio}.
\newblock \bibinfo{journal}{\emph{IEEE Transactions on Audio, Speech and
  Language Processing}}  \bibinfo{volume}{33} (\bibinfo{year}{2025}),
  \bibinfo{pages}{386--400}.
\newblock


\bibitem[Xue et~al\mbox{.}(2026a)]%
        {xue2026rtcfake}
\bibfield{author}{\bibinfo{person}{Jun Xue}, \bibinfo{person}{Zhuolin Yi},
  \bibinfo{person}{Yihuan Huang}, \bibinfo{person}{Yanzhen Ren},
  \bibinfo{person}{Yujie Chen}, \bibinfo{person}{Cunhang Fan},
  \bibinfo{person}{Zicheng Su}, \bibinfo{person}{Yongcheng Zhang}, {and}
  \bibinfo{person}{Bo Cai}.} \bibinfo{year}{2026}\natexlab{a}.
\newblock \showarticletitle{Rtcfake: Speech deepfake detection in real-time
  communication}. In \bibinfo{booktitle}{\emph{Findings of the Association for
  Computational Linguistics: ACL 2026}}. \bibinfo{pages}{5763--5775}.
\newblock


\bibitem[Xue et~al\mbox{.}(2026b)]%
        {xue2026profiling}
\bibfield{author}{\bibinfo{person}{Jun Xue}, \bibinfo{person}{Tong Zhang},
  \bibinfo{person}{Zhuolin Yi}, \bibinfo{person}{Yihuan Huang},
  \bibinfo{person}{Yi Chai}, \bibinfo{person}{Yiyang Zhang}, {and}
  \bibinfo{person}{Yanzhen Ren}.} \bibinfo{year}{2026}\natexlab{b}.
\newblock \showarticletitle{Profiling the Voice: Speaker-Specific Phoneme
  Fingerprinting for Speech Deepfake Detection}.
\newblock \bibinfo{journal}{\emph{arXiv preprint arXiv:2605.17737}}
  (\bibinfo{year}{2026}).
\newblock


\bibitem[Yi et~al\mbox{.}(2022)]%
        {yi2022add}
\bibfield{author}{\bibinfo{person}{Jiangyan Yi}, \bibinfo{person}{Ruibo Fu},
  \bibinfo{person}{Jianhua Tao}, \bibinfo{person}{Shuai Nie},
  \bibinfo{person}{Haoxin Ma}, \bibinfo{person}{Chenglong Wang},
  \bibinfo{person}{Tao Wang}, \bibinfo{person}{Zhengkun Tian},
  \bibinfo{person}{Ye Bai}, \bibinfo{person}{Cunhang Fan}, {et~al\mbox{.}}}
  \bibinfo{year}{2022}\natexlab{}.
\newblock \showarticletitle{Add 2022: the first audio deep synthesis detection
  challenge}. In \bibinfo{booktitle}{\emph{ICASSP 2022-2022 IEEE International
  Conference on Acoustics, Speech and Signal Processing (ICASSP)}}. IEEE,
  \bibinfo{pages}{9216--9220}.
\newblock


\bibitem[Yi et~al\mbox{.}(2023)]%
        {yi2023add}
\bibfield{author}{\bibinfo{person}{Jiangyan Yi}, \bibinfo{person}{Jianhua Tao},
  \bibinfo{person}{Ruibo Fu}, \bibinfo{person}{Xinrui Yan},
  \bibinfo{person}{Chenglong Wang}, \bibinfo{person}{Tao Wang},
  \bibinfo{person}{Chu~Yuan Zhang}, \bibinfo{person}{Xiaohui Zhang},
  \bibinfo{person}{Yan Zhao}, \bibinfo{person}{Yong Ren}, {et~al\mbox{.}}}
  \bibinfo{year}{2023}\natexlab{}.
\newblock \showarticletitle{Add 2023: the second audio deepfake detection
  challenge}.
\newblock \bibinfo{journal}{\emph{arXiv preprint arXiv:2305.13774}}
  (\bibinfo{year}{2023}).
\newblock


\bibitem[Zhang et~al\mbox{.}(2024)]%
        {zhang2024audio}
\bibfield{author}{\bibinfo{person}{Qishan Zhang}, \bibinfo{person}{Shuangbing
  Wen}, {and} \bibinfo{person}{Tao Hu}.} \bibinfo{year}{2024}\natexlab{}.
\newblock \showarticletitle{Audio deepfake detection with self-supervised xls-r
  and sls classifier}. In \bibinfo{booktitle}{\emph{Proceedings of the 32nd ACM
  International Conference on Multimedia}}. \bibinfo{pages}{6765--6773}.
\newblock


\bibitem[Zhou et~al\mbox{.}(2025)]%
        {zhou2025voxcpm}
\bibfield{author}{\bibinfo{person}{Yixuan Zhou}, \bibinfo{person}{Guoyang
  Zeng}, \bibinfo{person}{Xin Liu}, \bibinfo{person}{Xiang Li},
  \bibinfo{person}{Renjie Yu}, \bibinfo{person}{Ziyang Wang},
  \bibinfo{person}{Runchuan Ye}, \bibinfo{person}{Weiyue Sun},
  \bibinfo{person}{Jiancheng Gui}, \bibinfo{person}{Kehan Li}, {et~al\mbox{.}}}
  \bibinfo{year}{2025}\natexlab{}.
\newblock \showarticletitle{VoxCPM: Tokenizer-Free TTS for Context-Aware Speech
  Generation and True-to-Life Voice Cloning}.
\newblock \bibinfo{journal}{\emph{arXiv preprint arXiv:2509.24650}}
  (\bibinfo{year}{2025}).
\newblock


\end{thebibliography}

\end{document}